\def\ptitle{Study of a class of non-polynomial oscillator potentials}
\nopagenumbers
\hsize 6.0 true in 
\hoffset 0.25 true in 
\emergencystretch=0.6 in                 
\vfuzz 0.4 in                            
\hfuzz  0.4 in                           
\vglue 0.1true in
\mathsurround=2pt                        
\topskip=24pt                            
\def\nl{\noindent}                       
\def\np{\hfil\vfil\break}                
\def\ppl#1{{\leftskip=10cm\noindent #1\smallskip}} 
\def\title#1{\bigskip\noindent\bf #1 ~ \tr\smallskip} 

\font\tr=cmr10                          
\font\bf=cmbx10                         
\font\it=cmti10                         
\font\trbig=cmbx10 scaled 1500          
\font\tiny=cmr8                        
\def\bra{{\rm <}}    
\def\ket{{\rm >}}    
\def\nl{\hfil\break\noindent}  
\def\np{\hfil\vfil\break}
\def\ppl#1{{\noindent\leftskip 9 cm #1\vskip 0 pt}} 

\def\bra{{\rm <}} 
\def\ket{{\rm >}} 
\def\hi#1#2{$#1$\kern -2pt-#2} 
\def\hy#1#2{#1-\kern -2pt$#2$} 

\def\dbox#1{\hbox{\vrule 
\vbox{\hrule \vskip #1\hbox{\hskip #1\vbox{\hsize=#1}\hskip #1}\vskip #1
\hrule}\vrule}}
\def\qed{\hfill \dbox{0.05true in}} 

\def\vr{\vrule height 12 true pt depth 6 true pt}
  
\def\vrq{\vr\quad} 

\def\ma#1{\hbox{\vrule #1}}             
\def\mb#1{\hbox{\bf#1}}                 
\def\ng{>\kern -9pt|\kern 9pt}          
\def\bra{{\rm <}}                       
\def\ket{{\rm >}}                       
\def\hi#1#2{$#1$\kern -2pt-#2}          
\def\hy#1#2{#1-\kern -2pt$#2$}          

\def\frac#1#2{{{#1}\over{#2}}}
\def\dbox#1{\hbox{\vrule  
                        \vbox{\hrule \vskip #1
                             \hbox{\hskip #1
                                 \vbox{\hsize=#1}%
                              \hskip #1}%
                         \vskip #1 \hrule}%
                      \vrule}}
\def\qed{\hfill \dbox{0.05true in}}  
\output={\shipout\vbox{\makeheadline
                                      \ifnum\the\pageno>1 {\hrule}  \fi 
                                      {\pagebody}   
                                      \makefootline}
                   \advancepageno}
\headline{\noindent {\ifnum\the\pageno>1 
                                   {\tiny \ptitle\hfil page~\the\pageno}\fi}}
\footline{}
\newcount\zz  \zz=0  
\newcount\q   
\newcount\qq    \qq=0  
\def\pref #1#2#3#4#5{\frenchspacing \global \advance \q by 1     
    \edef#1{\the\q}
       {\ifnum \zz=1 { %
         \item{[\the\q]} 
         {#2} {\bf #3},{ #4.}{~#5}\medskip} \fi}}
\def\bref #1#2#3#4#5{\frenchspacing \global \advance \q by 1     
    \edef#1{\the\q}
    {\ifnum \zz=1 { %
       \item{[\the\q]} 
       {#2}, {\it #3} {(#4).}{~#5}\medskip} \fi}}
\def\gref #1#2{\frenchspacing \global \advance \q by 1  
    \edef#1{\the\q}
    {\ifnum \zz=1 { %
       \item{[\the\q]} 
       {#2}\medskip} \fi}}
 \def\sref #1{~[#1]}
 
\def\references#1{\zz=#1
   \parskip=2pt plus 1pt   
   {\ifnum \zz=1 {\noindent \bf References \medskip} \fi} \q=\qq

\pref{\bi}{Biswas S N, Saxen R P, Srivastava P K and Varma V S, J. Math. Phys. }{14}{1190 (1973)}{}
\pref{\ss}{Salam A and Strathdee J, Phys. Rev. D }{1}{3296 (1970)}{}
\pref{\ri}{Risken H and Vollmer H D, Z. Phys. }{201}{323 (1967)}{}
\bref{\ha}{Haken H }{Laser Theory}{ Encyclopedia of Physics XXV/2c,  Princeton  NJ: Van Nostrand, 1970}{}
\pref{\mi}{Mitra A K, J. Math. Phys. }{19}{2018 (1978)}{}
\pref{\ka}{Kaushal S R, J. Phys. A: Math. Gen. }{12}{L253 (1979)}{}
\pref{\ki}{Killingbeck J, Comput. Phys. Commun. }{18}{211 (1979)}{}
\pref{\gk}{Galicia S and Killingbeck J, Phys. Lett. A }{71}{17 (1979)}{}
\pref{\bb}{Bessis N and Bessis G, J. Math. Phys. }{21}{2780 (1980)}{}
\pref{\ho}{Hautot A, J. Comput. Phys. }{39}{72 (1981)}{}
\pref{\li}{Lai C S and Lin H E, J. Phys. A: Math. Gen. }{15}{1495 (1982)}{}
\pref{\wa}{Witwit M R M, Indian J. Pure \& Appl. Phys.  }{32}{391 (1994)}{}
\pref{\bh}{Bessis N, Bessis G and Hadinger G, J. Phys. A: Math. Gen. }{16}{497 (1983)}{}
\pref{\co}{Cohen M, J. Phys. A: Math. Gen. }{17}{2345 (1984)}{}
\pref{\fv}{Fack V and Vanden Berghe G, J. Phys. A: Math. Gen. }{18}{3355 (1985)}{}
\pref{\fb}{Fack V and Vanden Berghe G, Comp. Phys. Commun. }{39}{187 (1986)}{}
\pref{\fmv}{Fack V, De Meyer H and Vanden Berghe G, J. Math. Phys. }{27}{1340 (1986)}{}
\pref{\fvb}{Fack V and Vanden Berghe G, J. Phys. A: Math. Gen. }{20}{4153 (1987)}{}
\pref{\hn}{Handy C R, J. Phys. A: Math. Gen. }{18}{2141 (1985)}{}
\pref{\hw}{Hodgson R J W, J. Phys. A: Math. Gen. }{21}{1563 (1988)}{}
\pref{\av}{Agrawal R K and Varma V S, Phys. Rev. A }{48}{1921 (1993)}{}
\pref{\fs}{Flessas G P, Phys. lett. A }{83}{121 (1981)}{}
\pref{\va}{Varma V S, J. Phys. A: Math. Gen. }{14}{L489 (1981)}{}
\pref{\wh}{Whitehead R P, Watt A, Flessas G P and Nagarajan M A, J. Phys. A: Math. Gen. }{15}{1217 (1982)}{}
\pref{\haa}{Heading J,  J. Phys. A: Math. Gen. }{15}{2355 (1982)}{}
\pref{\hb}{Heading J,  J. Phys. A: Math. Gen. }{16}{2121 (1983)}{}
\pref{\fe}{Flessas G P, J. Phys. A: Math. Gen. }{15}{L97 (1982)}{}
\pref{\cm}{Chaudhuri R N and Mukherjee B, J. Phys. A: Math. Gen. }{16}{4031 (1983)}{}
\pref{\ma}{Marcilhacy G and Pons R, J. Phys. A: Math. Gen. }{18}{2441 (1985)}{}
\pref{\mb}{Marcilhacy G and Pons R, Phys. lett. A }{152}{235 (1991)}{}
\pref{\bl}{Blecher M H and Leach P G, J. Phys. A: Math. Gen. }{20}{5927 (1987)}{}
\pref{\ga}{Jason A C Gallas, J. Phys. A: Math. Gen. }{21}{3393 (1988)}{}
\pref{\rra}{Roy P and Roychoudhury R, Phys. lett. A }{122}{275 (1987)}{}
\pref{\rr}{Roy P, Roychoudhury R and Varshni Y P, J. Phys. A: Math. Gen. }{21}{1589 (1988)}{}
\pref{\vb}{Vanden Berghe G and De Meyer H E, J. Phys. A: Math. Gen. }{22}{1705 (1989)}{}
\pref{\lk}{Lakhtakia A, J. Phys. A: Math. Gen. }{22}{1701 (1989)}{}
\pref{\pr}{Roy P and Roychoudhury R, J. Phys. A: Math. Gen. }{23}{1657 (1990)}{}
\pref{\fr}{Fernandez F M, Phys. lett. A }{160}{116 (1991)}{}
\pref{\bv}{Bose S K and Varma N, Phys. lett. A }{141}{141 (1989)}{}
\pref{\hl}{Hislop D, Wolfaardt M F and Leach P G L, J. Phys. A: Math. Gen. }{23}{L1109 (1990)}{}
\pref{\sg}{Stubbins C and Gornstein M, Phys. lett. A }{202}{34 (1995)}{}
\pref{\hz}{Richard L Hall and Wei Hua Zhou, Can. J. Phy. }{76}{31 (1998)}{}
\pref{\Ih}{Ishikawa H, J. Phys. A: Math. Gen. }{35}{4453 (2002)}{}
\pref{\hh}{Handy C R, Hayes H, Stephens D V, Joshua J, and Summerour S, J. Phys. A: Math. Gen. }{26}{2649 (1993)}{}
\pref{\za}{Znojil M, J. Phys. A: Math. Gen. }{16}{279 (1983)}{}
\pref{\zb}{Znojil M, J. Phys. A: Math. Gen. }{17}{3441 (1984)}{}
\pref{\vr}{Varshni Y P, Phys. Rev. A }{36}{3009 (1987)}{}
\pref{\rp}{Roy B, Roychoudhury R and Roy P, J. Phys. A: Math. Gen. }{21}{1579 (1988)}{}
\pref{\sr}{Scherrer H, Risken H and Leiber T, Phys. Rev. A}{38}{3949 (1988)}{}
\pref{\ad}{Adhikari R, Dutt R and Varshni, J. Math. Phys. }{32}{447 (1991)}{}
\pref{\wi}{Witwit M R M, J. Phys. A: Math. Gen. }{24}{5291 (1991)}{}
\pref{\wm}{Witwit M R M, J. Comput. Phys.}{129}{220 (1996)}{}
\pref{\mo}{Mustafa O and Odeh M, Conference on Analysis and Mathematical Physics, Lund University, }{arXiv.org:math-ph/0101029}{(2001)}{}
\pref{\hs}{ Hall L R, Saad N and Keviczky A B, J. Math. Phys. }{43}{94 (2002)}{} 
\pref{\fl}{Ferreira C, L\'opez J L and Sinusía E P, Advances Appl. Math. }{34}{467 (2005)}{} 
\bref{\tn}{Temme M N}{Special Functions}{Wiley-Interscience Publication, New York (1996)}{Ch. 11, formula 11.7.}
\bref{\chen}{Heli Chen}{the quadrature discretization method and its applications}{Ph. D. thesis, University of British Columbia, 1998}{}
\pref{\heli}{Heli Chen and Bernie D. Shizgal, J. Math. Chem.  }{24}{321 (1998)}{}
\pref{\ips}{Imbo T, Pagnamenta A and Sukhatme U, Phys. Rev. D }{29}{1669 (1984)}{}
}
 \references{0}    

\ppl{CUQM-115}
\ppl{math-ph/0605056}
\hskip 2 cm 

\tr 

\baselineskip = 15true pt  
\vskip 0.5 true in
\centerline{\bf\trbig \ptitle}
\medskip
\vskip 0.25 true in
\centerline{Nasser Saad$^\dagger$, Richard L. Hall$^\ddagger$ and Hakan Ciftci$^*$}
\vskip 0.25 true in
{\leftskip=0pt plus 1fil
\rightskip=0pt plus 1fil\parfillskip=0pt\baselineskip 10 pt
\obeylines
$^\dagger$ Department of Mathematics and Statistics
University of Prince Edward Island
550 University Avenue, Charlottetown
PEI, Canada C1A 4P3.\par}
\medskip
{\leftskip=0pt plus 1fil
\rightskip=0pt plus 1fil\parfillskip=0pt\baselineskip 10 pt
\obeylines
$^\ddagger$ Department of Mathematics and Statistics, Concordia University
1455 de Maisonneuve Boulevard West, Montr\'eal
Qu\'ebec, Canada H3G 1M8.\par}
\medskip
{\leftskip=0pt plus 1fil \rightskip=0pt plus 1fil\parfillskip=0pt
\baselineskip 10 pt
\obeylines $^*$ Gazi Universitesi, Fen-Edebiyat Fak\"ultesi 
Fizik B\"ol\"um\"u, 06500 Teknikokullar 
Ankara, Turkey.
\par}
\medskip
\vskip 0.5 true in
\centerline{\bf Abstract}
\bigskip\noindent We develop a variational method to obtain accurate bounds for the eigenenergies of $H = -\Delta + V$ in arbitrary dimensions $N>1,$ where $V(r)$ is the nonpolynomial oscillator potential $V(r)=r^2+{\lambda r^2\over 1+gr^2},~ \lambda \in (-\infty,\infty), g>0$. The variational bounds are compared with results previously obtained in the literature. An infinite set of exact solutions is also obtained and used as a source of comparison eigenvalues. 
\vskip 1 true in

\np 
\title{1. Introduction}
\medskip
\noindent The following is a variational study of the eigenvalues of the Hamiltonian $H = \Delta + V(r),$ where the non-polynomial oscillator potential $V(r)$ is given in arbitrary dimensions by
$$V(r)=r^2+{\lambda r^2\over 1+gr^2},\quad~g>0, \lambda\in (-\infty,\infty).\eqno(1.1)$$
Various techniques may be found in the literature \sref{\bi-\mo} for the estimation of the spectrum corresponding to this potential.  Some of these techniques have been developed only for special values of $\lambda$ and $g$. The purpose of the present work is to introduce a unified variational technique to compute the eigenvalues of Schr\"odinger's equation with potential (1.1) in arbitrary dimension $N>1$ and for arbitrary angular quantum number $l=0,1,2,\dots$. The variational method presented here has the advantage of being valid for arbitrary values of $\lambda$, $g>0$ in arbitrary dimensional $N$. The article is organized as follows. In section 2, we introduce a variational technique based on the Gol'dman and Krivchenkov Hamiltonian as a solvable model. Thereafter, we use its eigensolutions to compute the matrix elements for the operators $r^2/(1+gr^2)$, $g>0$. Closed-form analytic expressions in terms of the incomplete gamma function are obtained for the matrix elements.
In section 3, application to the non-polynomial oscillator potential is discussed and we compare our results with those by other techniques avaliable in the literature. Some remarks regarding the perturbation expansions are also given in section 4. Our conclusions are presented in section 5. For comparison, we review, in the appendix, the exact solutions of the Schr\"odinger equation with the potential (1.1) valid for certain relations between the parameters $\lambda$ and $g$. These exact solutions are valid for $\lambda<0$ and can be seen as generalizations of the exact solutions in the one-dimensional case developed earlier in the literature. The purpose of exploring these exact solutions is to use them for testing the accuracy of the variational method we develop in section 2. 
\bigskip
\noindent{\bf 2. Variational approximation by means of an exactly solvable model}
\medskip
\noindent The radial Schr\"odinger equation in $N$-space dimension for the $n^{th}$ quantum state of the non-polynomial oscillator potential (1.1) can be written, with $\hbar/2m=1$, as
$$H\phi_{nl}^N(r)=\bigg(-{d^2\over dr^2}+Br^2+{(l+{1\over 2}(N-3))(l+{1\over 2}(N-1))\over r^2}+{\lambda r^2\over 1+gr^2}\bigg)\phi_{nl}^N(r)=E_{nl}^N\phi_{nl}^N(r),\eqno(2.1)$$
where $l=0$ for the $N=1$ case. Here the parameter $B$ is introduced to simplify the computation of the matrix elements of $H$;  it may be set equal to $1$ afterwards, without disturbing the computation. For $N>1$, $n$ refers to the radial quantum number (that is to say, $n-1$ is the number of radial nodes). For certain relations between the parameters $\lambda$ and $g$, exact solutions can be obtained \sref{\fs,\va,\bl,\ga,\vb,\lk,\fr,\bv}. These exact solutions are summarized in the appendix and are valid for arbitrary values of $\gamma=l+{1\over 2}(N-3), $ i.e. not restricted to certain integer values of $N$ and $l$ as reported in the literature.  They will be used to verify the numerical computations by the variational technique we are about to develop. Our variational approach is based on the introduction of a dummy variational parameter $A$ on which the full perturbation expansion does not depend (though the truncated expansion does). The optimal value of the  variational parameter is then determined by a minimization process. Let us write the Hamiltonian $H$ in (2.1) as 
$$H=H_0+\bigg({\lambda r^2\over 1+gr^2}-{A\over r^2}\bigg),\quad A\geq -{1\over 4}\eqno(2.2)$$
where
$$H_0=-{d^2\over dr^2}+Br^2+{(l+{1\over 2}(N-3))(l+{1\over 2}(N-1))+A\over r^2},~ l=0,1,2,\dots\eqno(2.3)$$
The unperturbed Hamiltonian $H_0$, known as the Gol'dman and Krivchenkov Hamiltonian, is one of the few systems that admit exact analytical solutions\sref{\hs}. It is the generalization of the familiar harmonic oscillator in 3-dimension $ -{d^2/dr^2}+Br^2+{l(l+1)/r^2}$, where the generalization lies in the parameter $A$ ranging over $[-{1\over 4},\infty)$ instead of only values determined by the angular momentum quantum numbers $l=0,1,2,\dots$. The energy spectrum of the Schr\"odinger equation $H_0\psi_{nl}^N={\epsilon}_{nl}^N\psi_{nl}^N$ is given, in terms of parameters $A$ and $B$, by
$${\epsilon}_{nl}^N=2\beta(2n+\zeta),~ n=0,1,2,\dots,\quad (\beta=\sqrt{B}, \zeta=1+\sqrt{A+(l+{N\over 2}-1)^2})\eqno(2.4)$$
where the {\it normalized} wavefunctions are given explicitly by 
$$
\psi_{nl}^N(r)=(-1)^n\sqrt{{2\beta^{\zeta}(\zeta)_n}\over n! \Gamma(\zeta)} r^{\zeta-{1\over 2}} e^{-{\beta\over 
2}r^2}{}_1F_1\bigg(\matrix{-n\cr
	\zeta\cr}\bigg|\beta r^2\bigg),\quad (n,~l=0,1,2,\dots, N\geq 1).\eqno(2.5)
$$
The confluent hypergeometric function ${}_1F_1$ is defined by means of the polynomial
$${}_1F_1\bigg(\matrix{-n\cr
	b\cr}\bigg|z\bigg)=\sum\limits_{k=0}^n {{(-n)_kz^k}\over {(b)_kk!}},\quad\hbox{($n$-degree polynomial in $z$)}\eqno(2.6)$$
where the shifted factorial $(a)_n$ is defined by
$$(a)_0=1,\quad (a)_n=a(a+1)(a+2)\dots (a+n-1), \quad{\rm for}\ n = 1,2,3,\dots\eqno(2.7)$$
The singular basis, consisting of the set of exact solutions of $H_0$, serves as a good starting point for the analysis of the non-polynomial oscillator Hamiltonian $H$ given in (2.1). Indeed, the use of the basis (2.5) will allow us to overcome the difficulties which arise if one uses Hermite functions \sref{\sg}. The matrix elements of the Hamiltonian $H$, (2.1), in terms of the orthonormal basis (2.5), read
$$\bra\psi_{ml}^N|H|\psi_{nl}^N\ket=2\bigg(2n+\zeta\bigg)\delta_{mn}+
\bra\psi_{ml}^N|{\lambda r^2\over 1+gr^2}|\psi_{nl}^N\ket-\bra\psi_{ml}^N|Ar^{-2}|\psi_{nl}^N\ket,\eqno(2.8)$$
where the Kronecker delta $\delta_{mn}=0$ for $m\neq n$ and $1$ for $m=n$. Consequently, to make use of the matrix elements (2.8), we need to evaluate the expectation values  $\bra\psi_{ml}^N|{r^2\over 1+gr^2}|\psi_{nl}^N\ket$ and $\bra\psi_{ml}^N|r^{-2}|\psi_{nl}^N\ket$ by means of the orthonormal basis (2.5). This is sunnarized by the following two theorems. 
\medskip
\noindent{\bf Theorem 1:}~{\it By means of the orthonormal basis $\{\psi_{nl}^N\}_{n=0}^\infty$, where $\psi_{nl}^N(r)$ given by (2.5), we have 
$$\eqalign{ <\psi_{ml}^N|{r^2\over 1+gr^2}|\psi_{nl}^N>\equiv&\int\limits_0^\infty {r^2\psi_{ml}^N(r)\psi_{nl}^N(r)\over 1+gr^2}dr=(-1)^{m+n}{\beta^{\zeta}e^{\beta/ g}\over g^{\zeta+1}\Gamma(\zeta)}\sqrt{(\zeta)_n(\zeta)_m\over n!m!}\cr
&\times \sum\limits_{i=0}^n\sum\limits_{j=0}^m{(-n)_i(-m)_j\over (\zeta)_i(\zeta)_ji!j!}\bigg({\beta\over g}\bigg)^{i+j}\Gamma(\zeta+i+j+1)\Gamma(-\zeta-i-j;{\beta\over g}).}\eqno(2.9)$$
Here $\Gamma(-\zeta-i-j;{\beta\over g})$ is the incomplete gamma function defined by
$\Gamma(a,z)=\int_z^\infty t^{a-1}e^{-t}dt.$}
\medskip
\noindent{\bf Proof:} Direct integration of $\int_0^\infty {r^{2}\psi_{nl}^N(r)\psi_{nl}^N(r)\over 1+gr^2}dr$ using (2.5) yields
$$\eqalign{\int\limits_0^\infty {r^2\psi_{ml}^N(r)\psi_{nl}^N(r)\over 1+gr^2}&dr={2(-1)^{m+n}\beta^{\zeta}\over \Gamma(\zeta)}\sqrt{(\zeta)_n(\zeta)_m\over n!m!}\int_0^\infty {r^{2\zeta+1}e^{-\beta r^2}\over 1+gr^2}
{}_1F_1\bigg(\matrix{-n\cr
	\zeta\cr}\bigg|\beta r^2\bigg){}_1F_1\bigg(\matrix{-m\cr
	\zeta\cr}\bigg|\beta r^2\bigg)dr\cr
&={2(-1)^{m+n}\beta^{\zeta}\over \Gamma(\zeta)}\sqrt{(\zeta)_n(\zeta)_m\over n!m!}
\sum\limits_{i=0}^n\sum\limits_{j=0}^m{(-n)_i(-m)_j\beta^{i+j}\over (\zeta)_i(\zeta)_ji!j!}\int\limits_0^\infty {r^{2\zeta+2i+2j+1}\over 1+gr^2}e^{-\beta r^2}dr}\eqno(2.10)
$$ where we have used the terminated series representation of the confluent hypergeometric function (2.6). The question is now reduced to the computation of the finite integral
$$I(\beta)=\int\limits_0^\infty {r^{2\zeta+2i+2j+1}\over 1+gr^2}e^{-\beta r^2}dr={1\over 2}\int\limits_0^\infty {x^{\zeta+i+j}\over 1+gx}e^{-\beta x}dx,\quad \hbox{(after substituting  } r^2=x).\eqno(2.11)$$
Such integrals have been questioned in the literature\sref{\mi,\bb,\co,\cm,\sg}; therefore, we shall give a detail calculation of it here. Differentiation of (2.11) with respect to $\beta$ yields
$${dI(\beta)\over d\beta}=-{1\over 2}\int\limits_0^\infty {x^{\zeta+i+j+1}\over 1+gx}e^{-\beta x}dx\eqno(2.12)$$
Multiplying (2.12) by $g$ and subtracting it from (2.11) implies
$$I(\beta)-g{dI(\beta)\over d\beta}={1\over 2}\int\limits_0^\infty x^{\zeta+i+j}e^{-\beta x}dx$$
The integral on the right hand side can be easily calculated by using the definition of the gamma function $\Gamma(z)=\int_0^\infty e^{-t}t^{z-1}dt,~ Re(z)>0$, and yields
$$I(\beta)-g{dI(\beta)\over d\beta}={1\over 2}\beta^{-(\zeta+i+j+1)}\Gamma(\zeta+i+j+1).$$
This is an elementary first-order differential equation which can be easily solved as
$$I(\beta)={1\over e^{-{\beta\over g}}}\bigg[-{\Gamma(\zeta+i+j+1)\over 2g}\int_{\epsilon>0}^\beta e^{-{\eta\over g}}\eta^{-(\zeta+i+j+1)}d\eta+C\bigg],$$ 
where the constant $C$ is chosen such that $I(\beta)\rightarrow 0$ as $\beta\rightarrow \infty$ (a conclusion which follows from the definition of the Laplace transform). Therefore, we have, after simple algebra and use of L'Hospital's rule, that
$$I(\beta)={\Gamma(\zeta+i+j+1)\over 2ge^{-{\beta\over g}}}\int_{\beta}^\infty e^{-{\eta\over g}}\eta^{-(\zeta+i+j+1)}d\eta={\Gamma(\zeta+i+j+1)\over 2g^{\zeta+i+j+1}e^{-{\beta\over g}}}\int_{\beta\over g}^\infty e^{-\eta}\eta^{-(\zeta+i+j+1)}d\eta$$ 
which is the definition of the incomplete gamma function. This completes the proof.\qed
\medskip
\noindent{\bf Theorem 2:}~{\it By means of (2.5) , we have 
$$<\psi_{ml}^N|r^{-2}|\psi_{nl}^N>=\cases{(-1)^{m+n}
{\beta\over \zeta-1}
\sqrt{n!(\zeta)_m\over m!(\zeta)_n}& if $n> m$,\cr
\ \cr
{\beta\over \zeta-1}& if $n=m$,\cr
\ \cr
(-1)^{m+n}
{\beta\over \zeta-1}
\sqrt{m!(\zeta)_n\over n!(\zeta)_m}& if $n<m$.\cr}
\eqno(2.13)
$$
}
\medskip
\noindent{\bf Proof:} Direct integration of $\int_0^\infty r^{-2}\psi_{nl}^N(r)\psi_{nl}^N(r)dr$ using (2.5) yields
$$\bra\psi_{nl}^N(r)| r^{-2}|\psi_{nl}^N(r)\ket={2(-1)^{m+n}\beta^\zeta\over \Gamma(\zeta)}\sqrt{(\zeta)_n(\zeta)_m\over nm!}\int_0^\infty r^{2\zeta-3}e^{-\beta r^2}{}_1F_1(-n;\zeta;\beta r^2){}_1F_1(-m;\zeta;\beta r^2)dr.$$
The finite integral on the right-hand side follows directly by means of the identity\sref{\hs}
$$\int\limits_0^\infty r^{2\zeta-\alpha-1} e^{-\beta r^2}
{}_1F_1\bigg(\matrix{-n\cr
	\zeta\cr}\bigg|\beta r^2\bigg){}_1F_1\bigg(\matrix{-m\cr
	\zeta\cr}\bigg|\beta r^2\bigg)dr={({\alpha\over 2})_n\Gamma(\zeta-{\alpha\over 2})\over2\beta^{\zeta-{\alpha\over 2}}(\zeta)_n} {}_3F_2\bigg(\matrix{-m,{\zeta-{\alpha\over 2}},{1-{\alpha\over 2}}\cr
	\zeta,1-{\alpha\over 2}-n\cr}\bigg|1\bigg).\eqno(2.14)
$$
The proof of the theorem then follows from the fact that the hypergeometric function
$${}_3F_2\bigg(\matrix{-m,a,b\cr
	c,d\cr}\bigg|1\bigg)=\sum\limits_{k=0}^m{(-m)_k(a)_k(b)_k\over (c)_k(d)_k\ k!},\quad (m-\hbox{degree polynomial)}.$$
collapses to unity when $\alpha=2$.\qed

\noindent Theorems 1 and 2 provide closed-form analytic expressions for the matrix elements (2.8); these allow us to obtain accurate upper bounds for the Hamiltonian (2.1) by means of the following optimization problem: 
$$E_{nl}^N\leq \min_{A\geq-{1\over 4}}\ \hbox{diag}\pmatrix{H_{00}&H_{01}&\dots&H_{0D-1}\cr
		    H_{10}&H_{11}&\dots&H_{1D-1}\cr
		    \dots&\dots&\dots&\dots\cr
		    H_{D-10}&H_{D-11}&\dots&H_{D-1D-1}}, \quad \quad H_{mn}=\bra\psi_{ml}^N|H|\psi_{nl}^N\ket,\eqno(2.15)$$
for fixed $B$.
\medskip
\noindent {\bf 3. Applications and Numerical Results}
\medskip
\noindent In order to avoid the numerical complication\sref{\fl} of computing the incomplete gamma functions in (2.8), we have, by means of  the recurrence relation\sref{\tn}
$$\Gamma(a+1,z)=a\Gamma(a,z)+z^ae^{-z},\eqno(3.1)$$
that
$$\Gamma(a,z)={\Gamma(a+n,z)\over (a)_n}-e^{-z}\sum\limits_{k=1}^n {z^{a+k-1}\over (a)_k}.\eqno(3.2)$$
Thus
$$
\Gamma(-\zeta-i-j,{\beta\over g})={(-1)^{i+j}\Gamma(-\zeta,{\beta\over g})\over (1+\zeta)_{i+j}}-\big({g\over \beta}\big)^{\zeta+i+j+1}e^{-{\beta\over g}}\sum\limits_{k=1}^{i+j} {({\beta\over g})^{k}\over (-\zeta-i-j)_k}.\eqno(3.3)$$
Consequently, (2.9) now becomes
$$\eqalign{<\psi_{ml}^N|{r^2\over 1+gr^2}|\psi_{nl}^N>&=(-1)^{m+n}{\zeta\beta^{\zeta}e^{\beta/ g}\over g^{\zeta+1}}\sqrt{(\zeta)_n(\zeta)_m\over n!m!}\Gamma(-\zeta,{\beta\over g})\sum\limits_{i=0}^n\sum\limits_{j=0}^m{(-n)_i(-m)_j\over (\zeta)_i(\zeta)_ji!j!}\big(-{\beta\over g}\big)^{i+j}\cr
&-{(-1)^{m+n}\over \beta\Gamma(\zeta)}\sqrt{(\zeta)_n(\zeta)_m\over n!m!}\sum\limits_{i=0}^n\sum\limits_{j=0}^m\sum\limits_{k=1}^{i+j}{(-n)_i(-m)_j\Gamma(\zeta+i+j+1-k)\over (\zeta)_i(\zeta)_ji!j!}\big(-{\beta\over g}\big)^{k}
},\eqno(3.4)$$
where we have used the identity
$(-\zeta-i-j)_k={(-1)^k\over (\zeta+i+j+1)_{-k}}$. The double sum of the first term on the right-hand side of (3.4) can be written in terms of the confluent hypergeometric function (2.6). In this case, we have
$$\eqalign{<\psi_{ml}^N|{r^2\over 1+gr^2}|\psi_{nl}^N>&=(-1)^{m+n}{\zeta\beta^{\zeta}e^{\beta/ g}\over g^{\zeta+1}}\sqrt{(\zeta)_n(\zeta)_m\over n!m!}\Gamma(-\zeta,{\beta\over g}){}_1F_1\bigg(\matrix{-n\cr
	\zeta\cr}\bigg|-{\beta\over g}\bigg){}_1F_1\bigg(\matrix{-m\cr
	\zeta\cr}\bigg|-{\beta\over g}\bigg)\cr
&-{(-1)^{m+n}\over \beta\Gamma(\zeta)}\sqrt{(\zeta)_n(\zeta)_m\over n!m!}\sum\limits_{i=0}^n\sum\limits_{j=0}^m\sum\limits_{k=1}^{i+j}{(-n)_i(-m)_j\Gamma(\zeta+i+j+1-k)\over (\zeta)_i(\zeta)_ji!j!}\big(-{\beta\over g}\big)^{k}
}\eqno(3.5)$$
The advantage of using such expressions as (3.4) or (3.5) is that they allow us to compute the incomplete gamma function only once, since its parameters are fixed. In our numerical computations of the variational upper bounds, we shall use (3.4) which can be implemented using symbolic mathematical software such as Mathematica or Maple. In order to verify the usefulness of this variational approach, as well as its generality for computing upper bounds for arbitrary values of the parameters $\lambda$ and $g>0$, we will now compare our results with the exact eigenvalues obtain for certain relations of the parameters (as summarized in the appendix) and with the most recent approximate eigenvalues obtained in the literature by means of a variety of techniques.  
\medskip
\noindent {\bf 3.1 Comparison with exact solutions}
\medskip
\noindent We first compare the variational upper bounds with the exact solutions developed in the appendix. It is known\sref{\wh} that such exact solutions are only possible for $\lambda<0$ and for certain algebraic relations between the parameters $\lambda$ and $g$. For the Schr\"odinger equation (2.1) with $\gamma=l+{1\over 2}(N-3)$ or real number $\gamma \geq -1$ and $B=1$, we have using (A.4) for $n=1$ that 
$$H\psi=-{d^2\psi\over dr^2}+\bigg(r^2+{\gamma(\gamma+1)\over r^2}-{4g+2g^2(3+2\gamma)\over 1+gr^2}\bigg)\psi=(3+2\gamma)(1-2g)\psi\eqno(3.6)$$
with the exact nodeless ground state $\psi(r)=r^{\gamma+1}(1+gr^2)e^{-r^2/2}$. In this case, the variational method for computing the upper bound of $H$ in (3.6) converges rapidly for arbitrary values of $g>0$ and minimization of $2\times 2$ matrix (2.15) over $A$ is sufficient to obtain very accurate upper bounds, as indicated by the Table 1, in comparison with the exact eigenvalues given by (3.6).\medskip
\np 
\nl {\bf Table~1.}~~Comparison of the exact eigenvalues obtained by means of (3.6) and that obtain by the variational technique $E_{var}$ discussed in the present work. The number $D$ in the square brackets refers the size of the basis set used in this case.
$$\vbox{
\hrule
\settabs
\+ \kern 0.5 true in\ &\vrq \kern 0.3true in &\vrq \kern 0.5true in &\vrq \kern 1.3true in &\vr\cr\hrule\+\cr
\+ $g$& $\gamma$& $\lambda$& $E_{exact}$& $E_{var}$\cr\hrule\+\cr
\+ $0.1$& $0$& $-0.46$& $~~2.400~000~000~000$& $~~2.400~000~000~000~[2]$\cr
\+ $~$& $2$& $-0.54$& $~~5.600~000~000~000$& $~~5.600~000~000~000$~[2]\cr\+\cr\hrule\+\cr
\+ $1$& $0$& $-10.0$& $-3.000~000~000~000$& $-3.000~000~000~000$~[2]\cr
\+ $~$& $2$& $-18.0$& $-7.000~000~000~000$& $-7.000~000~000~000$~[2]\cr\+\cr\hrule\+\cr
\+ $10$& $0$& $-640$& $-57.000~000~000~000$& $-57.000~000~000~000$~[2]\cr
\+ $~$& $2$& $-1440$& $-133.000~000~000~000$& $-133.000~000~000~000$~[2]\cr\+\cr\hrule\+\cr
}$$
Further, from (A.4) for $n=2$, we note that for $\lambda=(E_{\pm}-2\gamma-11)g$ where
$$E_\pm=5+2\gamma-g(13+6\gamma)\pm \sqrt{g^2(7+2\gamma)^2+g(-4+8\gamma)+4},\eqno(3.7)$$
the correspondent eigenstates reads  
$$\psi_2^\pm(r)=r^{\gamma+1}(1+gr^2)(1+a_\pm r^2)e^{-r^2/2},\eqno(3.8)$$
where
$$
a_\pm=-{E_\pm+(2g-1)(3+2\gamma)\over (6+4\gamma)}.
$$
For $E_+$, we have $a_+<0$ and the wave function $\psi_2^+(r)$ has one node on $[0,\infty)$, thus  $\psi_2^+(r)$ is the eigenstate for the first excited state in this case. On the other hand, for $E_-$, $a_->0$ and the wave function  $\psi_2^-(r)$ has no node and thus represent the ground state wave function when $g$ and $\lambda$ satisfy the correspondence condition. In Table 2, we have compared our variational results with those obtained by (3.7).  It should be mentioned that these variational eigenvalues are obtained by optimizing (2.15) over a {\it two}-dimensional subspace. The best possible value of the free minimization parameter $A$ is found to be zero. 

\nl {\bf Table~2.}~~Comparison of the eigenvalues obtained by means of (3.7) and those obtained by the variational technique $E_{var}$ discussed in the present work. The number $D$ in the square brackets refers the size of the basis set used in this case.
$$\vbox{
\hrule
\settabs
\+ \kern 0.5 true in\ &\vrq \kern 0.3true in &\vrq \kern 1.3true in &\vrq \kern 1.3true in &\vr\cr\hrule\+\cr
\+ $g$& $\gamma$& $\lambda$& $E_{exact}$& $E_{var}$\cr\hrule\+\cr
\+ $0.1$& $0$& $-26$& $-15.000~000~000~000$& $-14.9999999999999$~[2]\cr
\+ $~$& $~$& $-12$& $-~1.000~000~000~000$& $-0.99999999999996$~[2]\cr
\+ $~$& $1$& $-.56174575578973345189$& $~~7.382542442102665$& $~~7.382542442102671$~[2]\cr
\+ $~$& $~$& $-1.0182542442102665481$& $~~2.817457557897335$& $~~2.817457557897341$~[2]\cr
\+\cr\hrule
}$$
\medskip
\np 
\noindent {\bf 3.2 Comparison with the Quadrature Discretization Method}

\noindent In Table 3, we compare our upper bounds with that of Chen and Shizgal\sref{\chen,\heli} obtained by means of the Quadrature Discretization Method (QDM). The QDM generally employs a discretized version of the Hamiltonian with respect to a set of points that correspond to the quadrature points associated with the chosen weight function. The distribution of the grid points is determined by the weight function, which controls the convergence of the eigenvalues and eigenfunctions. In  Table 3, we have compared our results with those of\sref{\chen,\heli} using the weight functions $\exp(-\alpha r^2)$ and $\exp(-r^2\sqrt{1+\lambda/(1+0.5g)})$, where $\alpha$ is chosen by the authors for the fastest convergence.  The underlined portion of each eigenvalue indicates the convergence to the number of significant figures obtained by \sref{\chen,\heli} and has been mentioned for the purpose of comparison, we may extend the convergence of the upper bounds obtained by the variational method by increasing the size of $D$ in (2.15). The number $D$ indicates the size of the digonalized matrix used in the diagonalization process or/and the number of quadrature points used in the application of the QDM. The table clearly indicate the faster convergence the eigenvalues by means of the variational method. 
\np 
\noindent{\bf Table~3.}~~Comparison of the convergence of the lowest two odd eigenvalues of the nonpolynomial oscillator Hamiltonian $-d^2/dr^2+r^2+ r^2/(1+gr^2)$  between the variational method proposed in the present paper and the quadrature discretization method (QDM)\sref{\chen} with different weight functions $w(r)$. The number $D$ is the square brackets refers the size of the basis set used in each method.

$$\vbox{
\hrule
\settabs
\+ \kern 0.5true in\ &\vrq \kern 0.5true in &\vrq \kern 1.3true in &\vrq \kern 1.3true in &\vrq \kern 1.5true in &\vr\cr\hrule\+\cr
\+ $g$& $n$&$E_{\rm var}$&$E_{\rm QDM}$&$E_{\rm QDM}$\cr
\+ $~$& $~$&$~$&$w(r)=e^{-\alpha r^2}$&$w(r)=e^{- r^2\sqrt{1+\lambda/(1+0.5g)}}$\cr\hrule\+\cr
\+ $1$& $1$& $3.51158094~~[D=1]$& $3.51099389~~[D=10]$&$3.50666367~[D=10]$\cr
\+ $~$& $~$&$3.50740682~~[D=5]$& $3.50738872~~[D=20]$&$3.50737573~[D=20]$\cr
\+ $~$& $~$&$3.50738865~~[D=10]$& $3.50738837~~[D=25]$&$3.50738781~[D=30]$\cr
\+ $~$& $~$&$3.50738836~~[D=15]$& $\underline{3.50738835}~~[D=30]$&$3.50738831~[D=40]$\cr
\+ $~$& $~$&$\underline{3.50738835}~~[N=16]$& $3.50738835~~[D=30]$&$3.50738835~[D=50]$\cr
\hrule\+\cr
\+ $~$& $2$&$7.65220193~~[D=1]$& $7.83801690~~[D=10]$&$7.64562064~[D=10]$\cr
\+ $~$& $~$&$7.64828294~~[D=5]$& $7.64836479~~[D=20]$&$7.64815336~[D=20]$\cr
\+ $~$& $~$&$7.64820250~~[D=10]$& $7.64820406~~[D=25]$&$7.64819920~[D=30]$\cr
\+ $~$& $~$&$7.64820129~~[D=15]$& $\underline{7.64820124}~~[D=30]$&$7.64820110~[D=40]$\cr
\+ $~$& $~$&$\underline{7.64820124}~~[D=19]$& $7.64820124~~[D=35]$&$7.64820124~[D=60]$\cr
\hrule\+\cr
\+ $10$& $1$& $3.08809337~~[D=1]$& $3.54168906~~[D=10]$&$3.08692234~[D=10]$\cr
\+ $~$& $~$&$3.08809139~~[D=5]$& $3.08883073~~[D=30]$&$3.08794408~[D=30]$\cr
\+ $~$& $~$&$3.08809096~~[D=10]$& $3.08809133~~[D=50]$&$3.08805698~[D=50]$\cr
\+ $~$& $~$&$3.08809088~~[D=15]$& $\underline{3.08809085}~~[D=60]$&$3.08809057~[D=150]$\cr
\+ $~$& $~$&$\underline{3.0880908}6~~[D=19]$& $3.08809085~~[D=70]$&$\underline{3.088090}75~[D=180]$\cr
\hrule\+\cr
\+ $~$& $2$&$7.09037623~~[D=1]$& $11.03978681~~[D=10]$&$7.08838586~[D=10]$\cr
\+ $~$& $~$&$7.09037144~~[D=5]$& $7.13612019~~[D=30]$&$7.09012096~[D=20]$\cr
\+ $~$& $~$&$7.09037060~~[D=10]$& $7.09048160~~[D=50]$&$7.09031284~[D=50]$\cr
\+ $~$& $~$&$7.09037046~~[D=15]$& $7.09037053~~[D=70]$&$7.09036993~[D=150]$\cr
\+ $~$& $~$&$\underline{7.0903704}3~~[D=19]$&$\underline{7.09037041}~~[D=80]$&$\underline{7.090370}25~[D=180]$\cr
\hrule\+\cr
\+ $100$& $1$& $3.009831772~~[D=1]$& $6.61289499~~[D=10]$&$3.00981139~[D=10]$\cr
\+ $~$& $~$&$3.009831771~~[D=5]$& $3.00987806~~[D=100]$&$3.00982785~[D=100]$\cr
\+ $~$& $~$&$3.009831771~~[D=10]$& $3.00983181~~[D=150]$&$3.00983006~[D=170]$\cr
\+ $~$& $~$&$\underline{3.009831771}~~[D=15]$& $\underline{3.00983177}~~[D=170]$&$\underline{3.00983}023~[D=180]$\cr
\hrule\+\cr
\+ $~$& $2$&$7.00984496~~[D=1]$& $25.73416848~~[D=10]$&$7.00981397~[D=10]$\cr
\+ $~$& $~$&$7.00984495~~[D=5]$& $7.01481472~~[D=30]$&$7.00983900~[D=100]$\cr
\+ $~$& $~$&$7.00984495~~[D=10]$& $7.00985617~~[D=50]$&$7.00984172~[D=150]$\cr
\+ $~$& $~$&$\underline{7.00984495}~~[D=15]$&$\underline{7.009845}17~~[D=180]$&$\underline{7.00984}261~[D=180]$\cr
\hrule\+\cr
}$$
\np 
\noindent {\bf 3.3 Comparison with the Eigenvalue Moment Method}
\medskip
\noindent The eigenvalue moment method (EMM) is an effective and simple technique for  generating converging lower and upper bounds to the low-lying discrete spectrum of multidimensional quantum Hamiltonians\sref{\hn,\hh}. The method, however, faced some difficulties in determining eigenenergies for certain Hamiltonian parameter values. For the non-polynomial oscillator potential (1.1), we have compared our variational results obtained by means of (3.15) with those of EMM \sref{\hh}. As seen from the table, EMM is not adequate for large values of $g$, whereas the variational method developed here gives good results.
\np
\nl {\bf Table~4.}~~Comparison of the variational upper bounds $E_{var}$ with the upper and lower bounds $E_{EMM}$ obatin by means of the the eigenvalue moment method (EMM)\sref{\hh}. The number $D$ in the square brackets refers the size of the basis set used in each method. The size of the basis set used in the variational method is set to be $D=18$.
$$\vbox{
\hrule
\settabs
\+ \kern 0.4true in\ &\vrq \kern 0.3true in &\vrq \kern 0.3true in &\vrq \kern 1.3true in &\vrq \kern 1.5true in &\vr\cr\hrule\+\cr
\+ $\lambda$& $g$&$l$& $E_{var}$& $E_{EMM}$\cr\+\cr\hrule\+\cr
\+ $0.1$& $0.1$& 1& $5.186373002931507$& $5.1863730029314<E<5.1863730029316$\cr
\+ $~$& $~$&$2$& $7.243961840421887$& $7.2439618404138<E<7.2439618404260$\cr
\+ $~$& $~$&$3$& $9.294359110874627$& $9.29435911086337<E<9.29435911088159$\cr\+\cr\hrule\+\cr
\+ $0.1$& $0.5$& 1& $5.100857624300696$& $5.100842<E<5.100865$\cr
\+ $~$& $~$& 2& $7.11898087156427$& $7.11890<E<7.118901$\cr
\+ $~$& $~$& 3& $9.131812401691521$& $9.131799<E<9.131838$\cr\+\cr\hrule\+\cr
\+ $0.1$& $1$& 1& $5.065569521783354$& $5.06428<E<5.06609$\cr
\+ $~$& $~$& 2& $7.073726361909647$& $7.0730<E<7.0744$\cr
\+ $~$& $~$& 3& $9.078911720303639$& $9.0787<E<9.07892$\cr\+\cr\hrule\+\cr
\+ $0.5$& $1$& 1& $5.893595152339402$& $5.89359515233919<E<5.89359515233945$\cr
\+ $~$& $~$& 2& $8.1778716934677$& $8.177871693435<E<8.177871693485$\cr
\+ $~$& $~$& 3& $10.429204118147453$& $10.4292041181366<E<10.4292041181548$\cr\+\cr\hrule\+\cr
\+ $1$& $0.1$& 1& $6.704238892478644$& $6.7042388924777<E<6.7042388924788$\cr
\+ $~$& $~$& 2& $9.261914780826569$& $9.2619147807<E<9.2619147809$\cr
\+ $~$& $~$& 3& $11.760620962669972$& $11.7606209626312<E<11.7606209626917$\cr\+\cr\hrule\+\cr
\+ $1$& $1$& 1& $5.65139331725017$& $5.6503<E<5.6521$\cr
\+ $~$& $~$& 2& $7.73482804292358$& $7.734<E<7.736$\cr
\+ $~$& $~$& 3& $9.787669778509466$& $9.7875<E<9.7881$\cr\+\cr\hrule\+\cr
\+ $100$& $100$& 1& $5.993438873366758$& $-<E<6.389$\cr
\+ $~$& $~$& 2& $7.996024673021835$& $7.9947<E<8.037800$\cr
\+ $~$& $~$& 3& $9.997153638602487$& $9.9969<E<10.0113$\cr\+\cr\hrule\+\cr
}$$
\np 
\noindent {\bf 3.4 Some remarks about first-order perturbation expansions}
\medskip
\noindent The closed-form expressions for the matrix-elements (2.9) can be used explicitly to express the first-order perturbation expansions for the perturbed Hamiltonian (2.1) for any dimension $N$. Indeed, we have by means of (2.9) that
$$E_{nl}^N=2\beta(2n+\zeta)+
{\lambda \beta^\zeta e^{\beta/g}\over g^{\zeta+1}\Gamma(\zeta)}{(\zeta)_n\over n!}
\sum\limits_{i=0}^n\sum\limits_{j=0}^n{(-n)_i(-n)_j\over (\zeta)_i(\zeta)_ji!j!}\bigg({\beta\over g}\bigg)^{i+j}\Gamma(\zeta+i+j+1)\Gamma(-\zeta-i-j;{\beta\over g})-{A\beta\over \zeta-1}
$$
In particular, if $A=0$ and $B=1$, we have
$$E_{nl}^N=4n+N+2l+
{\lambda e^{1/g}({N\over 2}+l)_n\over n!g^{{N\over 2}+l+1}\Gamma({N\over 2}+l)}
\sum\limits_{i=0}^n\sum\limits_{j=0}^n{(-n)_i(-n)_jg^{-i-j}\over ({N\over 2}+l)_i({N\over 2}+l)_ji!j!}\Gamma({N\over 2}+l+i+j+1)\Gamma(-{N\over 2}-l-i-j;{1\over g}).
$$
In Table 5, we have expressed the first few states for different values of $g$ and for small values of $\lambda$ in the case of $N=3$ and $l=0$. 

\nl {\bf Table~5.}~~The first-order perturbation expansions of the lowest three eigenvalues for different values $g$ and valid for small values of $\lambda$. For comparison see Ref.\sref{\cm}.
$$\vbox{
\hrule
\settabs
\+ \kern 0.5 true in\ &\vrq \kern 1.3true in &\vrq \kern 1.3true in &\vrq \kern 1.3true in &\vr\cr\hrule\+\cr
\+ $g$& $E_{00}^3-3$& $E_{10}^3-7$& $E_{20}^3-11$\cr\hrule\+\cr
\+ $0.5$& $0.741\ 907\ 668\ 608\ 869\lambda$& $1.058~912~626~973~532\lambda$& $1.216~304~645~477~702\lambda$\cr
\+ $1$& $0.515~744~312~282~624\lambda$& $0.648~934~634~510~945\lambda$& $0.712~059~158~182~293\lambda$\cr
\+ $2$& $0.327~839~771~209~399\lambda$& $0.374~239~389~891~732\lambda$& $0.396~830~711~146~285\lambda$\cr
\+ $5$& $0.160~824~698~125~120\lambda$& $0.169~855~585~054~398\lambda$& $0.174~677~060~833~675\lambda$\cr
\+ $10$& $0.088~111~301~584~084\lambda$& $0.090~376~621~370~169\lambda$& $0.091~681~154~216~976\lambda$\cr
\+ $20$& $0.046~566~260~901~426\lambda$& $0.047~083~627~877~117\lambda$& $0.047~401~822~090~664\lambda$\cr
\+ $100$& $0.009~831~778~572~526\lambda$& $0.009~844~958~882~145\lambda$& $0.009~853~945~311~642\lambda$\cr
\+ $500$& $0.001~992~603~359~200\lambda$& $0.001~992~880~765~846\lambda$& $0.001~993~079~888~610\lambda$\cr
\+\cr\hrule
}$$
\medskip
\noindent {\bf 4. Conclusion}
\medskip
\noindent We have developed and applied a variational method to compute accurate upper bounds for the energies corresponding to the non-polynomial oscillator potential (1.1) in arbitrary dimension $N.$ The results show the effectiveness of this approach in comparison with other techniques. We have compared our variational results with those of many other techniques developed in the literature, for example with the Pseudo-Perturbation Expansion  Method\sref{\mo}, the Shifted $1/N$ Expansion\sref{\vr}, and the matrix continued fraction algorithm\sref{\sr}. For each technique, we have confirmed their reults with the use of very few matrix elements. Indeed, {\it all} of our variational bounds were obtained by the optimization of a square matrix (2.15) with size $D$ not exceeding $20$; these results can, of course, be further improved by simply increasing $D$.
\np 
\title{Acknowledgments}
\medskip
\noindent Partial financial support of this work under Grant Nos. GP3438 and GP249507 from the 
Natural Sciences and Engineering Research Council of Canada is gratefully 
acknowledged by two of us (respectively [RLH] and [NS]).

\bigskip
\title{Appendix: An infinite set of exact solutions}
\medskip
\noindent In this appendix we investigate the exact solutions of the non-polynomial oscillator Schr\"odinger equation in the $N$-dimenional case. The existence of an infinite number of exact solutions of the one-dimensional Schr\"odinger equation (2.1) ($B=1, N=1,l=0$) with eigenfunctions given by products of exponential and polynomial functions of $r^2$, for specific relations between the couplings $g$ and $\lambda$, was first reported by Flessas\sref{\fs} and soon afterwards was extended by Varma\sref{\va}. The main point of exploring these exact solutions in this appendix is that: (1) they generalize the previous exact solutions obtained for the one-dimensional case and they are valid for arbitrary $\gamma=l+{1\over 2}(N-3) \geq -1$; (2) they provide interesting examples for testing the vadility and accuracy of our variational technique. Assume 
$$\psi(r)\equiv \phi_{nl}^N(r)=r^{\gamma+1} e^{-r^2/2}\sum_{i=0}^n \alpha_i r^{2i}\eqno(A.1)$$
and substitute it into equation (2.1) yields ($B=1$, $E\equiv E_{nl}^N$)
$$\eqalign{&(E-2\gamma-3)\alpha_0+(6+4\gamma)\alpha_1+\sum_{i=1}^{n-1}\bigg\{\bigg(-2g(2i-2)-\lambda+g(E-2\gamma-3)\bigg)\alpha_{i-1}\cr
&+\bigg(2ig(2i-2)-4i+4ig(\gamma+1)+E-2\gamma-3\bigg)\alpha_i+\bigg(2(i+1)(2i+2\gamma+3)\bigg)\alpha_{i+1}\bigg\}r^{2i}\cr
&+\bigg\{\bigg(-4g(n-1)-\lambda+g(E-2\gamma-3)\bigg)\alpha_{n-1}+\bigg(2ng(2n-1)-4n+4ng(\gamma+1)+E-2\gamma-3\bigg)\alpha_n\bigg\}r^{2n}\cr
&+
\bigg(g(E-2\gamma-3)-\lambda-4gn\bigg)a_n r^{2n+2}=0}$$
The last term of this equation implies 
$$\lambda=g(E-4n-2\gamma-3),\quad n=0,1,2,\dots\eqno(A.2)$$
provided that the energy $E$ and the interaction parameters $\lambda,$ $g$ satisfies the $(n+1)\times (n+1)$-tridiagonal determinant
$${\cal D}_{n}(E)=\left|\matrix{b_0&a_1&0&0&\dots&0&0&0\cr
c_0&b_1&a_2 &0&\dots&0&0&0\cr
0&c_1&b_2&a_3&\dots&0&0&0\cr
\dots&\dots&\dots&\dots&\dots&\dots&\dots&\dots\cr
0&0&0&0&\dots&c_{n-2}&b_{n-1}&a_{n}\cr
0&0&0&0&\dots&0&c_{n-1}&b_{n}\cr
}\right|=0,\eqno(A.3a)$$
where 
$$
\cases{a_n= g(E-4n-2\gamma-3)-\lambda,&~\cr\cr
	b_n = 2ng(2n+2\gamma+1)-4n+E-2\gamma-3=E-4n-2\gamma-3+gc_n,&~\cr\cr
	c_n=2n(2n+2\gamma+1),\ n=0,1,2,\dots&~\cr}\eqno(A.3b)
$$
This determinantal equation can be written using the LU-decomposition as 
$$\eqalign{{\cal D}_{n}(E)&=\left|\matrix{1&0&0&0&\dots&0&0&0\cr
t_0&1&0&0&\dots&0&0&0\cr
0&t_1&1&0&\dots&0&0&0\cr
\dots&\dots&\dots&\dots&\dots&\dots&\dots&\dots\cr
0&0&0&0&\dots&t_{n-2}&1&0\cr
0&0&0&0&\dots&0&t_{n-1}&1\cr
}\right| \left|\matrix{u_0&v_1&0&0&\dots&0&0&0\cr
0&u_1&v_2 &0&\dots&0&0&0\cr
0&0&u_2&v_3&\dots&0&0&0\cr
\dots&\dots&\dots&\dots&\dots&\dots&\dots&\dots\cr
0&0&0&0&\dots&0&u_{n-1}&v_{n}\cr
0&0&0&0&\dots&0&0&u_{n}\cr
}\right|\cr}$$
in which
$$u_n= b_n-{c_na_{n-1}\over u_{n-1}},\quad u_0=b_0,\quad v_n=a_n,\quad t_n={c_n\over u_{n-1}}.
$$
This immediately leads to the following simple expression of computing the polynomial conditions as
$${\cal D}_{n}(E)=\prod_{k=1}^{n} u_{k}=\prod_{k=1}^{n}\bigg( b_n-{c_na_{n-1}\over u_{n-1}}\bigg)=0.\eqno(A.4)$$
It should be clear that our relations (A.3)-(A.4) summarize the exact solutions not only in the $N$-space dimensional where $\gamma=l+{1\over 2}(N-3)$ but also for arbitrary real values of $\gamma\geq -1$. \bigskip
\references{1}

\end